\PassOptionsToPackage{usenames,dvipsnames,svgnames,table}{xcolor}
\documentclass[conference]{IEEEtran} \usepackage{textcomp} \usepackage[nocompress]{cite}
\usepackage[pdftex]{graphicx}
\usepackage[utf8x]{inputenc}
\usepackage[T1]{fontenc}
\usepackage{url}
\usepackage{xspace}
\usepackage{xcolor}
\usepackage{marvosym}
\usepackage[export]{adjustbox}
\usepackage{booktabs, tabularx, multirow, makecell, siunitx}
\usepackage{eso-pic}
\usepackage[most]{tcolorbox}

\newcolumntype{R}{>{\raggedleft\arraybackslash}X}

\usepackage{setspace}
\usepackage[margin=3pt,skip=3pt,belowskip=0pt,font={small,stretch=0.9}]{caption}

\usepackage{acronym}
\usepackage{amsmath,amssymb,amsfonts}
\usepackage[abbreviations]{foreign}
\usepackage[colorlinks=true,linkcolor=black,anchorcolor=black,citecolor=black,filecolor=black,menucolor=black,runcolor=black,urlcolor=black]{hyperref}
\usepackage[capitalise]{cleveref}
\usepackage{pifont}
\usepackage[caption=false]{subfig}

\usepackage{hyperxmp}
\hypersetup{
    pdfauthor={Jämes Ménétrey, Marcelo Pasin, Pascal Felber, Valerio Schiavoni},
    pdftitle={Twine: An Embedded Trusted Runtime for WebAssembly},
}

\newcommand{\twine}{\textsc{Twine}\xspace}
\newcommand{\sys}{\twine}
\newcommand{\polybench}{PolyBench/C\xspace}
\newcommand{\ipfs}{Intel protected file system\xspace}

\def\insertLklMemBepc{2.8}
\def\insertLklFileBepc{2.8}
\def\insertLklMemAepc{15.7}
\def\insertLklFileAepc{3.2}
\def\insertWamrMem{4.4}
\def\insertWamrFile{2.7}
\def\insertTwineMemBepc{7.2}
\def\insertTwineFileBepc{8.2}
\def\insertTwineMemAepc{13.0}
\def\insertTwineFileAepc{8.7}
\def\insertLklVsTwineMemAepc{1.035}

\def\readSeqLklMemBepc{2.0}
\def\readSeqLklFileBepc{1.3}
\def\readSeqLklMemAepc{7.5}
\def\readSeqLklFileAepc{2.9}
\def\readSeqWamrMem{3.3}
\def\readSeqWamrFile{2.2}
\def\readSeqTwineMemBepc{5.1}
\def\readSeqTwineFileBepc{4.4}
\def\readSeqTwineMemAepc{8.5}
\def\readSeqTwineFileAepc{9.8}

\def\readRandLklMemBepc{2.2}
\def\readRandLklFileBepc{19.1}
\def\readRandLklMemAepc{16.8}
\def\readRandLklFileAepc{21.6}
\def\readRandWamrMem{2.4}
\def\readRandWamrFile{1.9}
\def\readRandTwineMemBepc{4.1}
\def\readRandTwineFileBepc{18.5}
\def\readRandTwineMemAepc{17.9}
\def\readRandTwineFileAepc{20.1}

\def\readRandLklVsTwineFileBepc{1.031}
\def\readRandLklVsTwineFileAepc{1.074}
 \def\speedtestWamrMemToNativeRatio{4.1}
\def\speedtestWamrFileToNativeRatio{3.7}
\def\speedtestTwineMemToWamrRatio{1.7}
\def\speedtestTwineFileToWamrRatio{1.9}
\def\speedtestExpFourOneZeroTwineMemVsFileRatio{12.4}
\def\speedtestExpFourOneZeroSgxLklMemVsFileRatio{22.1}
 \def\compilationNativeAverage{23,350}
\def\compilationSgxAverage{288,774}
\def\compilationWasmBinAverage{4,329}
\def\compilationWasmSgxBinAverage{3,425}
\def\compilationWasmAverage{38,593}
\def\optimizationSgxLklDiskImageAverage{15,711}
\def\optimizationWasmAotAverage{52,944}
\def\launchTimeNativeAverage{2}
\def\launchTimeSgxAverage{6,119}
\def\launchTimeWasmAverage{70}
\def\launchTimeWasmSgxAverage{3,155}
\def\memoryNativeInMemoryAverage{192,822}
\def\memorySgxInMemoryAverage{77,310}
\def\memoryWasmInMemoryAverage{211,156}
\def\memoryWasmSgxInMemoryAverage{9,970}
\def\memorySgxEnclaveSize{261,120}
\def\memoryWasmSgxEnclaveSize{209,920}
\def\sizeNativeAverage{1,164}
\def\sizeSgxLklBinAverage{6,546}
\def\sizeWasmBinAverage{123}
\def\sizeWasmSgxBinAverage{30}
\def\sizeSgxLklEnclaveAverage{79,200}
\def\sizeWasmSgxEnclaveAverage{567}
\def\sizeSgxLklDiskImageAverage{247,552}
\def\sizeWasmAverage{1,155}
\def\sizeWasmAotAverage{3,707}
\def\launchTimeTwineVsSgx{1.939}
 \def\memsetRatio{50.1}
\def\ocallRatio{36.2}
\def\otherOperationsRatio{10.7}
\def\sqliteRatio{2.9}
\def\copyRatio{75.9}
\def\copyInvertedRatio{24.1}
\def\insertRatio{1.5}
\def\seqReadingRatio{2.5}
\def\randReadingRatio{4.1}
 
\makeatletter
\begin{document}

\title{\twine: An Embedded Trusted Runtime\\
for WebAssembly}

\author{\IEEEauthorblockN{Jämes Ménétrey}
\IEEEauthorblockA{\textit{University of Neuchâtel}\\
Switzerland\\
james.menetrey@unine.ch}
\and
\IEEEauthorblockN{Marcelo Pasin}
\IEEEauthorblockA{\textit{University of Neuchâtel}\\
Switzerland\\
marcelo.pasin@unine.ch}
\and
\IEEEauthorblockN{Pascal Felber}
\IEEEauthorblockA{\textit{University of Neuchâtel}\\
Switzerland\\
pascal.felber@unine.ch}
\and
\IEEEauthorblockN{Valerio Schiavoni}
\IEEEauthorblockA{\textit{University of Neuchâtel}\\
Switzerland\\
valerio.schiavoni@unine.ch}
}

\maketitle

\definecolor{yellowPaper}{HTML}{fff8ae}
\AddToShipoutPictureFG*{\AtTextUpperLeft{\begin{tcolorbox}[width=\textwidth,colback=yellowPaper,enhanced,frame hidden,sharp corners]  
        \centering\scriptsize
        \copyright~2021 IEEE. Personal use of this material is permitted. Permission from IEEE must be obtained for all other uses, in any current or future media, including reprinting/republishing this material for advertising or promotional purposes, creating new collective works, for resale or redistribution to servers or lists, or reuse of any copyrighted component of this work in other works.
        This is the author's version of the work. The definitive version has been published in the proceedings of the\\
        37th IEEE International Conference on Data Engineering (ICDE'21).
        \href{https://doi.org/10.1109/ICDE51399.2021.00025}{DOI: 10.1109/ICDE51399.2021.00025}
     \end{tcolorbox}   
  }}

\hypersetup{
    pdfcopyright={\copyright~2021 IEEE. Personal use of this material is permitted. Permission from IEEE must be obtained for all other uses, in any current or future media, including reprinting/republishing this material for advertising or promotional purposes, creating new collective works, for resale or redistribution to servers or lists, or reuse of any copyrighted component of this work in other works.
    This is the author's version of the work. The definitive version will be published in the proceedings of the 37th IEEE International Conference on Data Engineering (ICDE'21).}
}

\begin{abstract}

WebAssembly is an increasingly popular lightweight binary instruction format, which can be efficiently embedded and sandboxed.
Languages like C, C++, Rust, Go, and many others can be compiled into WebAssembly.
This paper describes \twine, a WebAssembly trusted runtime designed to execute unmodified, language-independent applications.
We leverage Intel SGX to build the runtime environment without dealing with language-specific, complex APIs.
While SGX hardware provides secure execution within the processor, \twine provides a secure, sandboxed software runtime nested within an SGX enclave, featuring a WebAssembly system interface (WASI) for compatibility with unmodified WebAssembly applications.
We evaluate \twine with a large set of general-purpose benchmarks and real-world applications. 
In particular, we used \twine to implement a secure, trusted version of SQLite, a well-known full-fledged embeddable database.
We believe that such a trusted database would be a reasonable component to build many larger application services.
Our evaluation shows that SQLite can be fully executed inside an SGX enclave via WebAssembly and existing system interface, with similar average performance overheads.
We estimate that the performance penalties measured are largely compensated by the additional security guarantees and its full compatibility with standard WebAssembly.
An in-depth analysis of our results indicates that performance can be greatly improved by modifying some of the underlying libraries.
We describe and implement one such modification in the paper, showing up to $\randReadingRatio\times$ speedup.
\twine is open-source, available at GitHub along with instructions to reproduce our experiments.

\end{abstract}

\section{Introduction}\label{sec:intro}

Trusted code execution is currently one of the major open challenges for distributed systems.
Data is a key asset for many companies and the ability to execute code and process data out of premises is a prerequisite for outsourcing computing tasks, either to large data centres in the cloud or to the edge of the network on thin clients and IoT devices.
Trusted execution environments (TEEs) such as Intel SGX~\cite{cryptoeprint:2016:086}, ARM TrustZone~\cite{pinto2019demystifying}, AMD SME/SEV~\cite{amd-sev} and RISC-V Keystone~\cite{10.1145/3342195.3387532} gathered much attention lately as they provide hardware support for secure code execution within special hardware constructs that are shielded from the outside world, including the operating system and privileged users.
Still, despite the many frameworks and runtime environments that have been developed recently, programming applications for TEEs remains a complex task.
Developers must generally use custom tools and APIs, and they are restricted to a few supported programming languages.

In this paper, we propose a trusted runtime that supports execution of unmodified applications compiled to WebAssembly (Wasm)~\cite{10.1145/3140587.3062363}, a portable binary-code format for executable programs originally designed for efficient execution within Web browsers.
Among its many benefits, Wasm is optimised for speed, can be efficiently embedded, sandboxed, and is considered secure~\cite{lehmann2020everything}.
The LLVM compiler toolchain, one of the most popular compilation infrastructure nowadays, natively supports Wasm as a standard compilation target.
Thanks to that, programs developed in languages such as C, C++, Rust, Swift, Go, C\#, D, Delphi, Fortran, Haskell, Julia, Objective-C, and many others, can already be used as input to produce Wasm executables.
Therefore, by supporting Wasm, one can provide a generic runtime environment without resorting to language-specific, dedicated APIs.
Furthermore, this approach completely abstracts the application from the underlying hardware and operating system (OS).
\begin{figure}[!t]
	\centering
	\includegraphics[scale=0.6]{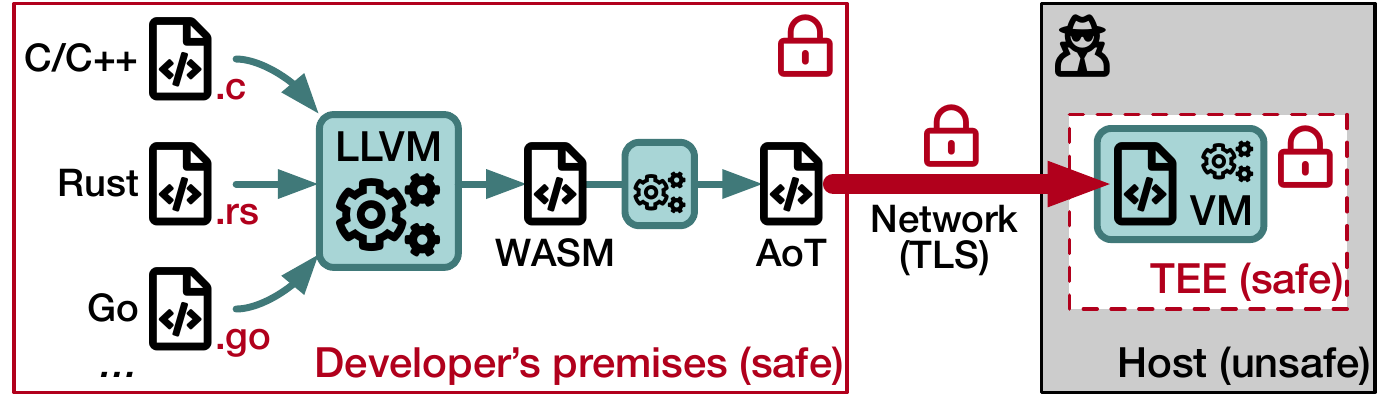}
	\caption{Overview of \twine's workflow.}
	\label{fig:overview}
\end{figure}

We present \twine (trusted Wasm in enclave), a lightweight embeddable Wasm virtual machine running in a TEE.
\Cref{fig:overview} depicts the typical \sys workflow.
It acts as an adaptation layer between the application and the underlying TEE, the OS and hardware.
\twine includes a comprehensive WASI (WebAssembly system interface) layer to allow for native execution of legacy Wasm applications, without recompilation.
We currently support Intel SGX enclaves as TEEs: \twine dynamically translates WASI operations into equivalent native OS calls or to functions from secure libraries purposely built for SGX.
In particular, \twine maps file operations to \ipfs~\cite{ipfs}, and persisted data is transparently encrypted and never accessible in plaintext from outside an enclave.
Whereas a TEE provides a secure hardware execution runtime in the processor, \twine provides a secure software runtime (sandbox) nested within the TEE, with a WASI interface for compatibility with legacy Wasm, abstracting the underlying environment from the application.

We evaluated \twine with several micro- and macro-benchmarks, as well as a full SQLite implementation.
We compared its performances against existing software packages, with and without secure operations inside a TEE.
Our results reveal that \twine performs on par with systems providing similar security guarantees.
We also observed non-negligible performance overheads due to execution within the TEE under some workloads.
We believe this penalty is largely compensated by the additional security guarantees and full compatibility with Wasm code thanks to the WASI interface.

The contributions presented in this paper are:
\begin{itemize}
	\item The first real open-source implementation of a general-purpose Wasm runtime environment within SGX enclaves with full support for encrypted file system operations;
	\item An extensive evaluation of our implementation, offering a good understanding of its performance costs and associated bottlenecks;
	\item A proposal for improving Intel protected file system, and a showcase of the derived performance improvements.
\end{itemize}
To present our contributions, we organised this paper into six sections.
In \S\ref{sec:relat} we survey related work.
We provide a background on Intel SGX and WebAssembly in \S\ref{sec:backg}.
The design and implementation details of \twine are described in \S\ref{sec:runti}.
We report on the thorough evaluation of our full prototype in \S\ref{sec:evalu}, and conclude in \S\ref{sec:concl}.
 \section{Related Work}\label{sec:relat}

We survey related work according to different criteria.
First, we look at systems with dedicated support for Wasm inside TEEs.
Then, we review proposals for generic TEE support for language runtimes.
Finally, given that our evaluation (\S\ref{sec:evalu}) shows how to use \twine with SQLite, we investigate alternative proposals to run data management systems inside Intel SGX, specifically databases with SQL support or simpler key-value store. 
To the best of our knowledge, \sys is the first system to seamlessly leverage Wasm and WASI to provide specific features of a TEE.

\textbf{WebAssembly and TEEs.}
AccTEE~\cite{10.1145/3361525.3361541} runs Wasm binaries inside Intel SGX enclaves, with the specific goal of implementing trustworthy resource accounting under malicious OSs.
It leverages the SGX-LKL~\cite{DBLP:journals/corr/abs-1908-11143} library OS to execute Wasm binaries inside SGX enclaves. 
AccTEE tracks resource usage directly inside the enclave (\eg, by counting Wasm instructions, memory allocation, I/O operations, \etc).
Their two-way sandbox (firstly from disjoint memory spaces for Wasm modules, and secondly from SGX itself) is similar to \twine's double-sandboxing approach.
AccTEE lacks support for IPFS, used by \twine to persist data and code.

Se-Lambda~\cite{10.1007/978-3-030-01701-9_25} is a library built on top of OpenLambda to deploy serverless programs over Function-as-a-Service (FaaS) platforms with the additional security guarantees of Intel SGX.
Se-Lambda shields the FaaS gateway inside enclaves, as well as the code of the functions deployed by the users, providing anti-tampering and integrity guarantees.
Besides, it protects attacks with a privileged monitoring module that intercepts and checks system call return values.
We believe similar defense mechanisms could be easily integrated into \sys.

Enarx~\cite{enarx} is an open-source project whose long-term goal is to allow for the execution of language-agnostic binaries into SGX enclaves, as well as other target TEEs.
It leverages Wasm to offer attestation and delivery of the applications.
Since it is at an early-stage development (currently missing several required features), we could not conduct an in-depth comparison and postpone this as part of future work.

\textbf{Embedding language runtimes in TEEs.}
There have been many efforts to embed other language runtimes into TEEs~\cite{mesapy,tsai2020civet,goltzsche2017trustjs,wang2019running}.
\sys deploys a lightweight and versatile Wasm runtime inside an SGX enclave, which is able to execute Wasm applications that have been compiled ahead-of-time for maximum performance.
Additionally, we developed a WASI layer to enable any compliant application to run inside our runtime seamlessly.

\textbf{Database systems and TEEs.}
Deployment and execution of full-fledged database systems inside restricted environments as TEEs are challenging tasks. 
CryptSQLite~\cite{wang2019cryptsqlite} executes SQLite inside SGX enclaves, protecting both the confidentiality and integrity of user data, with an average overhead of 21\% for SQL statements when compared to SQLite with symmetric encryption mechanisms enabled.
There are also SQLite drop-in alternatives~\cite{sqlitecrypt} relying on symmetric encryption schemes or specific security extensions. However, once data is fetched and decrypted for processing in memory, it is readable in clear by an attacker (\eg, a malicious system administrator, a compromised OS, a tainted memory-dumping process, \etc) with or without physical access to the machine. 
These attacks are prevented by our approach (and similar ones) relying on the encrypted memory regions handled transparently by the SGX shielding mechanisms.

EnclaveDB~\cite{priebe2018enclavedb} is a secure version of Hekaton (Microsoft SQL server’s in-memory database engine).
Tables, indexes and other metadata are placed into SGX enclaves, with support for integrity and freshness of the database log.
Queries are signed, encrypted and deployed via a trusted channel to an enclave running over a (possibly untrusted) database server. 
By running a complete Wasm binary (in our SQLite scenarios), pre-compiled queries, as well as the query compiler and optimiser, are executed inside SGX enclaves.
Always Encrypted~\cite{10.1145/3318464.3386141} extends Microsoft SQL server to keep data always encrypted (except for data inside TEEs). 
It fully supports Windows virtualization-based security (VBS)~\cite{winvbs} enclaves and partially SGX.
This scheme is designed to be applied only on the most privacy-sensitive columns, rather than the whole database, as \sys does.

StealthDB~\cite{vinayagamurthy2019stealthdb} runs on top of Postgres and leverages SGX enclaves, using extended (encrypted) data types.
It requires changes to the DBMS code, but with the advantage of limiting the performance overheads.
Table schemas and entries, as well as user queries are encrypted, being only decrypted inside enclaves.
\sys allows running unmodified applications without resorting to changes to the source code.

\section{Background}\label{sec:backg}

This section provides background information on Intel SGX in (\S\ref{sec:sgx}) and the Wasm ecosystem (\S\ref{sec:wasm}) to help understand the architecture and design of \sys. 

\subsection{Intel SGX}\label{sec:sgx}

Software Guard Extensions (SGX)~\cite{cryptoeprint:2016:086} are a set of processor instructions found in modern Intel processors~\cite{intel-sgx-procs} that allow programmers to create encrypted regions of memory, called \emph{enclaves}.
Enclave memory content is automatically encrypted and decrypted when read and written by instructions running inside the enclave itself.
Enclave encryption keys are kept inside the processor and no instruction has access to the keys, not even when running with high hardware privilege levels, as OSs and virtual machine managers do.
The memory inside an enclave is protected from any unauthorised access, even from machine administrators with physical access.

Enclave memory access is accelerated by using a large cache memory, called EPC (enclave page cache).
EPC size is limited, with the latest CPUs offering up to 256 MiB.
The processor keeps unencrypted copies of all enclave pages in EPC, and paging is used when the EPC is full.
The hardware also maintains cryptographic hashes for all enclave pages in EPC, in such a way that a modification from outside an enclave can be automatically detected.
The EPC helps reduce access time to encrypted memory but also limits the number of pages concurrently available.
Swapping degrades performance and enclaved applications should strive to avoid it~\cite{10.1145/3064176.3064219}.

Instructions inside enclaves can access data outside the enclave, but calling instructions outside requires a special \emph{out call} instruction (\texttt{OCALL}).
Upon an \texttt{OCALL}, the CPU exits the protected enclave to execute code on the outside.
Conversely, there is an \emph{enclave call} (\texttt{ECALL}) instruction to call code inside an enclave.
\texttt{OCALL} and \texttt{ECALL} instructions are slow because switching the context from inside to outside an enclave is costly (up to 13'100 CPU cycles in latest server-grade processors).
It has been shown that enclaved applications can avoid such calls to reduce performance loss~\cite{10.1145/3268935.3268942}.

In order to build composed software using enclaves, one must have a method to establish trust.
For example, a client must know if it can trust a given server and vice versa.
Intel SGX offers a remote attestation mechanism to prove that an enclave can be trusted.
Each processor has a secret key fused in its die, used to derive many other keys.
One of the derived keys is used to build enclave attestations, calculated as a signature of the whole contents of an enclave at its creation.
An external attestation service confirms that a given enclave runs a particular piece of code on a genuine Intel SGX processor, which can then be trusted by another component.

\subsection{WebAssembly}\label{sec:wasm}

WebAssembly (Wasm) is a W3C recommended open standard for a portable and executable binary code format.
It was originally designed to improve the performance of applications embedded in Web browsers, similar to the now-deprecated Microsoft ActiveX, and directly superseding \texttt{asm.js}~\cite{eich2015asm}.
Since then, its support was extended to standalone environments (\ie, outside browsers).
Full application execution, especially in standalone environments, requires access to OS services, \eg, process and memory management or I/O, typically available via common system calls (for instance, exposed by a POSIX interface).
Hence, the interaction of Wasm with the underlying OS is standardised through a specific API called WebAssembly system interface (WASI)~\cite{Mozilla2019StandardizingWASI}.
This interface allows for several implementations suited to different OSs and incorporating several non-functional abstractions, including virtualisation, sandboxing, access control, \etc.
In the latest specifications, the WASI interface consists of 45 functions covering various capabilities: access to process arguments and environment variables, file system interaction, events polling, process management, random number generation, socket interaction and time retrieval.

\newcommand{\YES}{\textcolor{NavyBlue}{\ding{51}}}
\newcommand{\NO}{\color{BrickRed}{\ding{55}}} \newcommand{\requireadaptation}{\textsuperscript{*}}
\newcommand{\theresnorust}{\textsuperscript{\textdagger}}
\begin{table}[!t]
\centering
\small
\setlength{\tabcolsep}{2pt}
\caption{\label{tab:wasmruntimes}Comparison of Wasm runtimes.}
\rowcolors{1}{gray!10}{gray!0}
  \begin{tabularx}{\columnwidth}{Xccccc}
  \toprule
  \rowcolor{gray!25}
  Wasm runtime                                  & Language        & Embeddable                  & Interpreter     & JIT       & AoT \\
  \midrule   
  Wasmtime~\cite{BytecodeAlliance000wasmtime}   & Rust            & \YES\requireadaptation\theresnorust  & \NO             & \YES      & \YES \\ Wasmer~\cite{wasmer} 	   				              & Rust            & \YES\requireadaptation\theresnorust  & \NO             & \YES      & \YES  \\
  Lucet~\cite{BytecodeAllianceLucet}		        & Rust            & \YES\requireadaptation\theresnorust  & \NO             & \NO       & \YES  \\
  WAVM~\cite{wavm}                              & C++           & \YES\requireadaptation      & \NO             & \YES      & \YES \\
  Wasm3~\cite{wasm3}	 					                & C               & \YES\requireadaptation      & \YES            & \NO       & \NO \\
  WAMR~\cite{wamr}							                & C               & \YES                        & \YES            & \YES      & \YES  \\
  \bottomrule
  \noalign{\vskip 2pt} 
  \multicolumn{6}{l}{\cellcolor{gray!0}\requireadaptation\scriptsize{Requires adaptation for SGX.} \hfill \cellcolor{gray!0}\theresnorust\scriptsize{Intel does not support Rust for enclave development.}}\\
  \end{tabularx}
\end{table}

There are currently several options to generate and execute Wasm code.
\emph{Emscripten}~\cite{10.1145/2048147.2048224} and \emph{Binaryen}~\cite{binaryen} can compile C/C++ into Wasm binaries with support for POSIX OS calls for standalone applications.
These tools can convert and execute legacy applications into their Wasm representation.
However, the conversion is only possible by requesting the Wasm runtime to expose functions that are generally bound to a specific OS, \ie, not a standard nor a public interface.
Wasm applications become tightly coupled to a given OS, defeating one of its main purposes, \ie, portability.
WASI solves the issue with a standard and lightweight interface that Wasm runtimes can comply with to support a large variety of interactions abstracted from the OS.
The introduction of this abstract layer limits the coupling of Wasm applications to just WASI.
As a result, Wasm applications using WASI are system-agnostic and can run on any compliant OS or browser.

LLVM~~\cite{DBLP:conf/cgo/LattnerA04} is a compilation toolchain for several different programming languages.
The compilation is split into front- and back-end modules.
The connection between them uses the LLVM intermediate representation code.
LLVM supports several front-end modules for various languages and, similarly, many back-ends to generate different binary formats.
Since v8.0, LLVM officially supports and can generate Wasm code with WASI. All compiler front-ends using recent LLVM versions can consequently generate Wasm code.
Note that, while Wasm represents an abstract machine, WASI represents its abstract OS, \ie, a standard interface to run Wasm applications outside of a browser.
Due to this tight dependency, tools generating Wasm code must be adapted to couple the Wasm code generated with WASI calls.

The execution of Wasm code must be handled by a dedicated runtime, able to execute the instructions and implementing WASI calls.
We discuss below the advantages and drawbacks of existing Wasm runtimes and explain why \sys settled for one of them.
Table~\ref{tab:wasmruntimes} summarises the main properties of the Wasm runtimes considered.
We compare them in terms of execution modes, implementation language and whether they can be embedded into a TEE, such as SGX enclaves.

Wasmtime~\cite{BytecodeAlliance000wasmtime} is a Rust-based standalone runtime.
It uses Cranelift~\cite{cranelift}, a low-level retargetable just-in-time (JIT) compiler with similarities to LLVM.
Wasmtime can be used by various programming languages thanks to the wrappers available with the runtime.
Embedding a JIT compiler inside an SGX enclave, despite its potential performance benefits, increases the trusted computing base by a large factor.
Moreover, Wasmtime and Cranelift are implemented in Rust: while tools exist to support Rust binaries in SGX enclaves~\cite{Wang2019Rust}, we opted in \twine for the well-supported standard Intel toolchain.

Lucet~\cite{BytecodeAllianceLucet} is a native Wasm compiler and runtime also implemented in Rust.
It is designed to safely execute untrusted WebAssembly programs embedded in third-party applications.
It supports ahead-of-time (AoT) compilation of Wasm applications using Cranelift.
While the runtime is not coupled to Cranelift as Wasmtime, Lucet presents similar integration challenges (Rust, large TCB).

Wasmer~\cite{wasmer} is a Rust-based Wasm runtime for lightweight and portable containers based on Wasm.
It allows for JIT and AoT compilations with multiple back-ends, including LLVM and Cranelift.
It supports the two prominent application binary interfaces (ABI): WASI and Emscripten.
We turned away from Wasmer for the same reason as the previous alternatives.

WAVM~\cite{wavm} is a Wasm virtual machine written in C++.
It supports both WASI and Emscripten ABIs and offers various extensions, such as 128-bit SIMD, thread management and exception handling.
While implemented in C++ , hence with native support for enclave development, its tight coupling with LLVM makes it difficult (if possible at all) to embed it inside an SGX enclave. 

Wasm3~\cite{wasm3} is a micro-interpreter for Wasm, optimised for size, able to execute in restricted memory environments and to provide fast startup latency.
It was designed for constrained edge devices with very limited resources (\eg, Arduino and Particle).
Having a reduced set of dependencies and small code base, it can easily fit within SGX enclaves.
However, it only supports interpreted code and, hence, provides limited performance for executing Wasm binaries. 

The WebAssembly micro runtime (WAMR)~\cite{wamr} is a standalone Wasm runtime supported by the \emph{bytecode alliance} open source community.
This runtime supports two interpreted execution modes, one slower and one faster, the former using less memory than the other.
It also supports two binary execution modes, AoT and JIT, both using LLVM.
WAMR is implemented in C with a small footprint (runtime binary size of 50\,KiB for AoT, 85\,KiB for interpreter) and very few external dependencies, which is ideal for small embedded devices with limited resources.
WAMR can be linked with SGX enclaves out of the box, which significantly simplifies the integration of Wasm and SGX.
We, therefore, opted for WAMR as underlying runtime for \twine, as detailed in \S\ref{sec:runti}.

\section{Trusted runtime for WebAssembly}\label{sec:runti}
\begin{figure}[!t]
	\centering
	\includegraphics[scale=0.6]{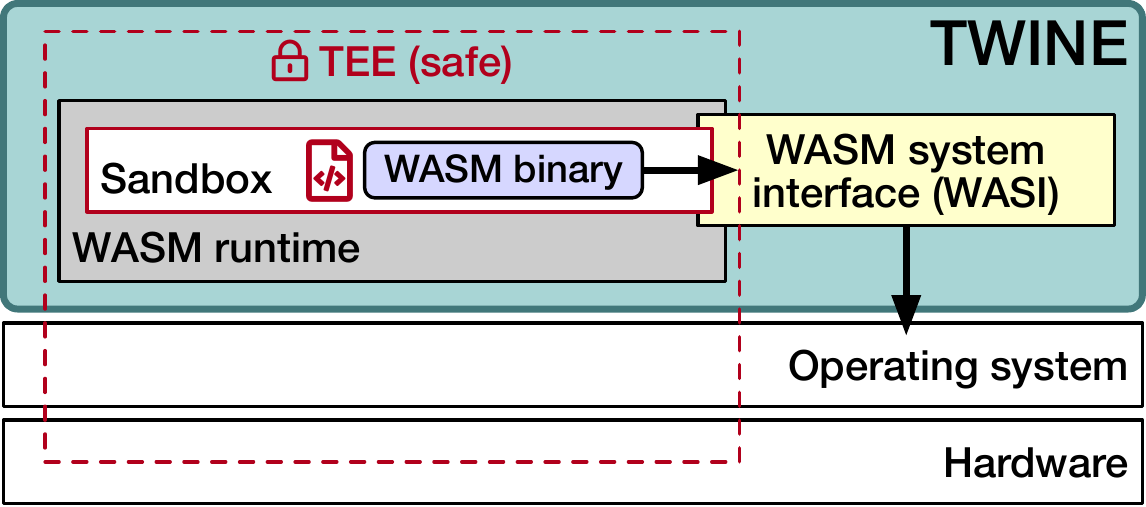}
	\caption{Overall \twine architecture.}
	\label{fig:arch}
\end{figure}

\twine is an execution environment suited for running Wasm applications inside TEEs.
It is built with two main blocks: a Wasm runtime and a WASI interface (see \Cref{fig:arch}).
The Wasm runtime runs entirely inside the TEE, and WASI works as a bridge between trusted and untrusted environments, abstracting the machinery dedicated to communicate with the underlying OS.
Thus, WASI is the equivalent to the traditional SGX adaptation layer comprised of the \texttt{OCALL}s.
The main advantage of relying on WASI is that it brings a triple abstraction.
Firstly, the programming language can be freely chosen by the developers, provided it can be compiled with LLVM or another compiler that supports Wasm and WASI as a compilation target.
This lifts the restrictions imposed by SGX, typically forcing enclaved applications to be written in C/C++.
Secondly, the TEE is abstracted away from the applications.
Applications can be safely executed as long as the TEE is able to interpret or execute Wasm (supported by WASI), opening the door to other TEE technologies.
Finally, WASI is system-agnostic, as long as the OS can provide an equivalent of the API required by WASI.
Since WASI mimics the system calls of POSIX systems, many Unix variants can implement it.

On top of its portability benefits, WASI focuses on security by sandboxing.
Regular applications usually call the OS through a standard interface (\eg, POSIX).
WASI adds a thin layer of control between Wasm OS calls and the actual OS interface.
As a result, the runtime environment can limit what Wasm can do on a program-by-program basis, preventing Wasm code from using the full rights of the user running the process.
For instance, a WASI implementation can restrict the application to a subtree of the file system, similar to the capabilities offered by \emph{chroot}.

The combination of the enclave and sandbox capabilities of SGX and WASI, respectively, ends up in a two-way sandboxing system partially inspired by MiniBox~\cite{minibox-atc14}.
The system, which is considered untrusted in the threat model of SGX, cannot compromise the integrity of the enclave code nor the confidentiality of the data stored in its memory.
Likewise, Wasm applications, considered untrusted from the system's owner standpoint, cannot interact directly with the OS unless WASI explicitly grants permission in the Wasm runtime.
Therefore, the Wasm application providers and the hosting platform can agree on the trust guarantees given by SGX and those of a reference \twine enclave with strong peer-reviewed sandboxing capabilities, making WASI a mutually trusted demilitarised zone.

\subsection{Threat model}

\twine leverages the protection of TEEs to offer a trusted environment for running Wasm applications.
Many guarantees offered by \twine are inherited from the underlying TEE, which in our implementation is Intel SGX.
Note that a different TEE may not withstand the same level of threats.

\textbf{Assumptions.}
We assume that no physical attack is possible against the computer hardware.
The TEE offers the level of protection as specified, and standard cryptography cannot be subverted.
Application and OS codes present no vulnerabilities by implementation mistake nor careless design.

\textbf{SGX enclaves.}
Code and data inside enclaves are considered as trusted, and nothing from outside can be considered trusted.
The non-enclaved part of a process, the OS and any hypervisor are thus potentially hostile.
The memory inside of an enclave can only be read in encrypted form from the outside.
Writing the memory enclave from the outside causes the enclave to be terminated.
Side-channel or denial-of-service attacks may exist, and applications running inside enclaves must be written to be resistant to them.
While we consider side-channel attacks out of scope, mitigations exist~\cite{10.1145/3359789.3359809,10.5555/3277355.3277378}.

\textbf{Operating system.}
The OS follows an honest-but-curious model.
In principle, the OS follows its specification and poses no threat to user processes.
A compromised OS may arbitrarily respond to enclave calls, causing its malfunction; enclaves should be carefully crafted to ignore abnormal responses or even abandon execution in such cases.

\subsection{WASI}

As presented in \S\ref{sec:backg}, we considered Wasmtime, Wasmer, Lucet, WAVM, Wasm3 and WAMR as runtime candidates for implementing \twine.
Wasmtime, Wasmer, Lucet and WAVM may be executed inside SGX enclaves, but require substantial adaptations to comply with the SGX enclaves' restrictions.
Moreover, some of these runtime environments (except WAVM and Wasm3) are written in Rust and require additional effort to use as a trusted runtime, since Intel does not support this programming language for enclave development.
Wasm3, on the other hand, is small but only offers an interpreter, this being an inadequate constraint for running standalone applications.
Finally, WAMR is also small, has few dependencies, and can link to binary code (albeit generated ahead of time, that is, no JIT).
We chose to use WAMR and replace its WASI interface, as explained below, in such a way that we can abstract the enclave constraints while implementing systems calls.

WASI is the interface through which Wasm applications communicate with the outside world, similar to POSIX's capabilities for regular native programs.
The development of TEE enabled applications requires to deal with crossing the boundary between trusted and untrusted environments, materialised with \texttt{ECALL}s and \texttt{OCALL}s in Intel SGX.
We believe that leveraging WASI as the communication layer meets the purpose of Wasm, where the implementation is abstracted away for the application itself.
As a result, the applications compiled in Wasm with WASI support do not require any modification to be executed inside a TEE.

The toolkit of WAMR provides an ahead-of-time compiler, enabling to compile Wasm applications into their native representation using LLVM before they reach \twine's enclave.
As such, \twine does not contain a Wasm interpreter and can only execute ahead-of-time compiled applications.
The main advantage of this choice is that native code execution is faster than code interpretation, which is critical to be competitive with the other secure TEE solutions~\cite{199364, 203255}.
Moreover, the Wasm runtime has a smaller memory footprint than the code interpreter, which are essential factors in the context of SGX and cloud/edge computing.
The option of embedding a JIT compiler was not considered, as bringing LLVM machinery in an enclave requires porting the code base to compile with the restrictions of SGX.

Unlike \twine, Intel SGX only guarantees the integrity of the enclave binary and not the confidentiality.
Integrity is verified with a signature in the code, but the code itself must be in plaintext to be loaded into an enclave memory.
\twine is able to offer the confidentiality of Wasm applications because the Wasm code is supplied using a secure channel after the enclave has been started.
When the Wasm code is received, it is mapped into a secure memory area called \emph{reserved memory}~\cite{IntelCorporation2020SGXDevRef}.
That memory area enables one to load arbitrary executable code and manage the pages' permissions as if they were outside the enclave.
Therefore, Wasm applications never leave the secure memory of the enclave.

\subsection{WASI implementation details}

By the time \twine was developed, WAMR already included a WASI implementation that relies heavily on POSIX calls.
POSIX is not available inside SGX enclaves, so the implementation of WASI written by the authors of WAMR needs to frequently cross the trusted boundary of the enclave and plainly routes most of the WASI functions to their POSIX equivalent using \texttt{OCALL}s.
While this approach enables to run any Wasm applications that comply with WASI inside an enclave, this does not bring additional security regarding the data that transits through POSIX.

We designed \twine to implement a different WASI interface for WAMR, that is more tailored to the specific TEE used (namely SGX).
We estimated that plainly forwarding WASI calls to outside the enclave was not the best option.
First, for performance reasons: most WASI calls would simply be translated to \texttt{OCALL}s.
Second, we wanted to leverage trusted implementations when available, as for instance \ipfs (IPFS), described below (\S\ref{sec:ipfs}).
Therefore, we refactored WAMR's WASI implementation to keep its sandboxing enforcement, and we split the remaining into two distinct layers, one for specific implementations, when available, and one for generic calls.
Generic calls are handled by calling a POSIX-like library outside the enclave while providing additional security measures and sanity checks.
Such calls are only implemented when no trusted compatible implementation exists.
For instance, time retrieval is not supported by Intel SGX.
Hence, \sys's POSIX layer leaves the enclave to fetch monotonic time while ensuring that the returned values are always greater than the previous ones.
If a trusted implementation exists (as the many in Intel SDK), we use it to handle its corresponding WASI call.
Sometimes a trusted implementation needs to call outside the enclave, but they often offer more guarantees than merely calling the OS.
One notable example is the protected file system, described below.
Finally, \sys includes a compilation flag to globally disable the untrusted POSIX implementation in the enclave, which is useful when developers require a strict and restricted environment or assess how their applications rely on external resources.
In particular, the interface may expose states from the TEE to the outside by leaking sensitive data in host calls, \eg, usage patterns and arguments, despite the returned values being checked once retrieved in the enclave.

Memory management greatly impacts on the performance of the code executed in an enclave (see \S\ref{sec:evalu}).
WAMR provides three modes to manage the memory for Wasm applications:
\emph{(1)}~the default memory allocator of the system,
\emph{(2)}~a custom memory allocator, and
\emph{(3)}~a buffer of memory.
\twine uses the latter option since we measured that an application that heavily relies on the memory allocator of SGX to enlarge existing buffers performs poorly.
For instance, SQLite micro-benchmarks in \S\ref{sec:microbenchmarks}, which requires to extend its internal buffer for every new record being added.
Before using a preallocated buffer for SQLite (see \S\ref{sec:envsetup}), we noticed the complexity of the SGX memory allocator to be above linear.

In its current implementation, \twine requires to expose a single \texttt{ECALL} to supply the Wasm application as an argument.
This function starts the Wasm runtime and executes the start routine of the Wasm application, as defined by WASI ABI specifications~\cite{wasiabi}.
Future versions of \twine would only receive the Wasm applications from trusted endpoints supplied by the applications providers, as shown in \Cref{fig:overview}.
The endpoint may either be hard-coded into the enclave code, and therefore part of the SGX measurement mechanism that prevents binary tampering, or provided in a manifest file with the enclave.
The endpoint can verify that the code running in the enclave is trusted using SGX's remote attestation.
As a result, \twine will provide both data and code confidentiality and integrity by relying on SGX capabilities, as well as a secure channel of communication between the enclave and the trusted application provider.
While the enclave must rely on the OS for network communication, the trusted code can use cryptographic techniques (\eg, elliptic-curve Diffie-Hellman) to create a channel that cannot be eavesdropped on.

\begin{figure*}[!t]
	\centering
	\includegraphics{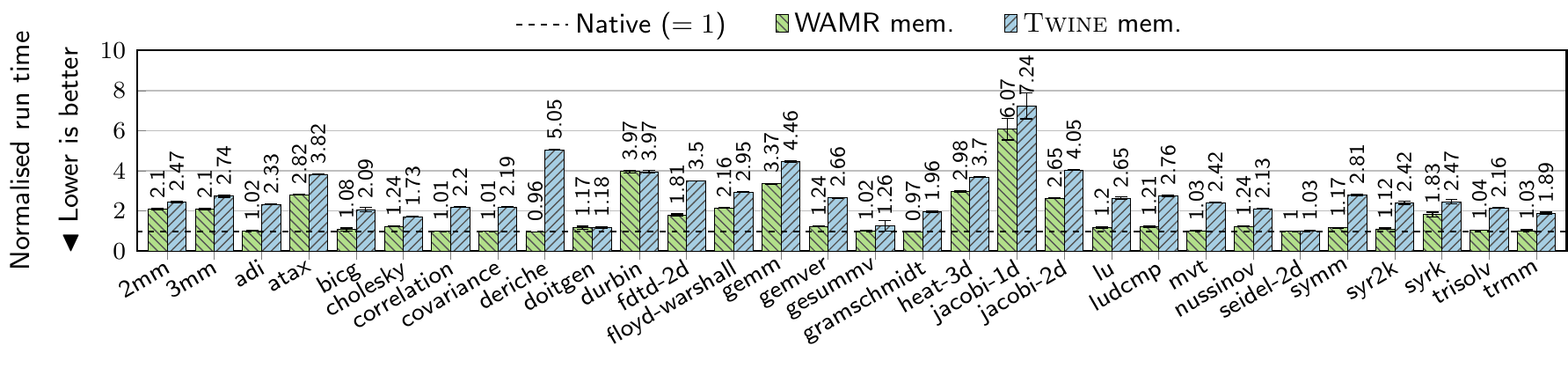}
	\vspace{-20pt}
	\caption{Performance of \polybench benchmarks, normalised to the native speed.}\label{fig:polybench}
	\vspace{-12pt}
\end{figure*}

\subsection{Intel Protected File System (IPFS)}\label{sec:ipfs}

To validate the abstraction offered by WASI, we implemented a subset of the WASI calls (\ie, those related to file system interaction) using the \ipfs~\cite{ipfs} (IPFS).
Part of Intel SGX SDK, it mimics POSIX standard functions for file input/output.
The architecture of IPFS is split in two: \emph{(1)} the trusted library, running in the enclave that offers a POSIX-like API for file management, and \emph{(2)} the untrusted library, an adapter layer to interact with the POSIX functions outside of the enclave, that actually read and write on the file system.
Upon a \texttt{write}, content is encrypted seamlessly by the trusted library, before being written on the media storage from the untrusted library.
Conversely, content is verified for integrity by the trusted enclave during reading operations.

IPFS uses AES-GCM for authenticated encryption, leveraging the CPU's native hardware acceleration. An encrypted file is structured as a Merkle tree with nodes of a fixed size of 4\,KiB. 
Each node contains the encryption key and tag for its children nodes.
Thus, IPFS iteratively decrypts parts of the tree as the program running in the enclave requests data~\cite{IPFSexplained}.
This mechanism ensures the confidentiality and the integrity of the data stored at rest on the untrusted file system.
While the enclave is running, the confidentiality and the integrity of the data are also guaranteed by SGX's memory shielding.

IPFS has several limitations, which are considered to be outside of its security objectives by Intel.
Since the files are saved in the regular file system, there is no protection against malicious file deletion and swapping.
Consequently, this technology lacks protection against:
\emph{(1)}~rollback attacks, IPFS cannot detect whether the latest version of the file is opened or has been swapped by an older version, and
\emph{(2)}~side-channel attacks, IPFS leak file usage patterns, and various metadata such as the file size (up to 4\,KiB granularity), access time and file name.
We note how Obliviate~\cite{ahmad2018obliviate}, a file system for SGX, partially mitigates such attacks.

\subsection{IPFS implementation details}

Many WASI API calls cannot be directly mapped to their equivalent functions in the IPFS, because Intel's implementation diverges from POSIX.
SQLite uses \texttt{fseek} to write data well after the end of a file, while IPFS' \texttt{sgx\_fseek} does not advance beyond the end of a file.
Our WASI implementation extends the file with null bytes, leveraging a few additional IPFS calls.
Also, IPFS lacks support for vectored read and write operations.
WASI function \texttt{fd\_read} is vectored, we therefore implemented it with an iteration.

IPFS provides convenient support to automatically create keys for encrypting files, derived from the enclave signature and the processor's (secret) keys.
While automatic key generation seems straightforward, a key generated by a specific enclave in a given processor cannot be regenerated elsewhere.
IPFS circumvents this limitation with a non-standard file open function, where the caller passes the key as a parameter.
Our prototype relies on automatic generation as an alternative to a trustworthy secret sharing service~\cite{9153433}.
We leave as future work to extend the SGX-enabled WASI layer to support encrypted communication through sockets.

In conclusion, files persisted by \twine are seen as ciphertext outside of the enclaves, while transparently decrypted and integrity-checked before being handled by a Wasm application.

\section{Evaluation}\label{sec:evalu}

We present here our extensive evaluation of \twine.
We intend to answer the following questions:
\begin{itemize}
	\item What is the performance overheads of using the runtime WAMR in SGX, compared to native applications?
	\item Can a database engine realistically be compiled into Wasm and executed in a TEE, while preserving acceptable performances? 
	\item How do the database input and output operations behave when the EPC size limit is reached?
	\item What are the primitives that generate most of the performance overheads while executing database queries? Can we improve them?
\end{itemize}

We answer these questions by using a general-purpose compute-bound evaluation with \polybench (\S\ref{sec:polybench}), evaluating a general-purpose embeddable database using SQLite (\S\ref{sec:speedtest1}), stressing the database engine using custom micro-benchmarks that perform read and write operations (\S\ref{sec:microbenchmarks}), analysing various cost factors bound to Wasm and SGX (\S\ref{sec:cost-factors}) and finally profiling the time breakdown of the database components, the Wasm runtime and the SDK of SGX (\S\ref{sec:profiling}).

\subsection{Experimental setup}\label{sec:envsetup}

We use a Supermicro SuperServer 5019S-M2, equipped with a 8-core Intel Xeon CPU E3-1275 v6 at 3.80GHz and 16GiB DDR4 2400 MHz.
We deploy Ubuntu 18.04.5 using kernel 4.15.0-128-generic, SGX driver v2.6.0, and the platform/SGX SDK v2.11.100.2.
The CPU supports SGX1, with an EPC limit set to 128\,MiB (usable 93\,MiB).

Time is measured using the POSIX function \texttt{clock} in all the benchmarks and averaged using the median.
If measured from within the enclave, the time to leave and reenter the enclave is included.
In our setup, the enclave round trip accounts for approximately 4\,ms.
We used Docker to build the benchmarks, while their execution is on bare metal to avoid potential isolation overheads.
The native benchmarks are compiled using Clang 10 with optimisation set to \texttt{--O3}.
The Wasm benchmarks are compiled using Clang into Wasm format, then AoT-compiled into native format using the compiler provided by WAMR (\ie, \texttt{wamrc}) using \texttt{--O3} and size level 1 to run into SGX enclaves (\texttt{--sgx}).
Finally, we used GCC v7.5.0 for two tasks: \emph{(1)} compile the applications executing the benchmarks, \ie, the WAMR runtime and the SGX enclaves, also with \texttt{--O3}, and \emph{(2)} compile IPFS with \texttt{--O2}, as in the SGX SDK.

SGX-LKL Open Enclave (v0.2.0) and LKL (v5.4.62) have been used as an empirical baseline for running the experiments natively in SGX enclaves.
They have been downloaded from the official Debian repository and compiled with optimisation \texttt{--O3}. 
Our implementation is open-source, and instructions to reproduce our experiments are available at GitHub~\cite{twine-github}.

\subsection{\polybench micro-benchmarks}\label{sec:polybench}

\polybench\cite{pouchet.11.polybench} is a CPU-bound benchmark suite commonly used to validate compiler optimisations and compare the performance of Wasm execution environments~\cite{DBLP:conf/usenix/JangdaPBG19,10.1145/3361525.3361541}.
We leveraged \polybench due to the practicality of deploying it in SGX enclaves.
We show the results for 30 \polybench (v4.2.1-beta) tests, compiled as native (plain x86-64 binaries) and Wasm compiled ahead-of-time.
Results are given for the native execution, those using WAMR for Wasm, and finally using \twine for Wasm in SGX.
\Cref{fig:polybench} shows the results normalised against the native run time.

\begin{figure*}[!t]
	\centering
	\includegraphics{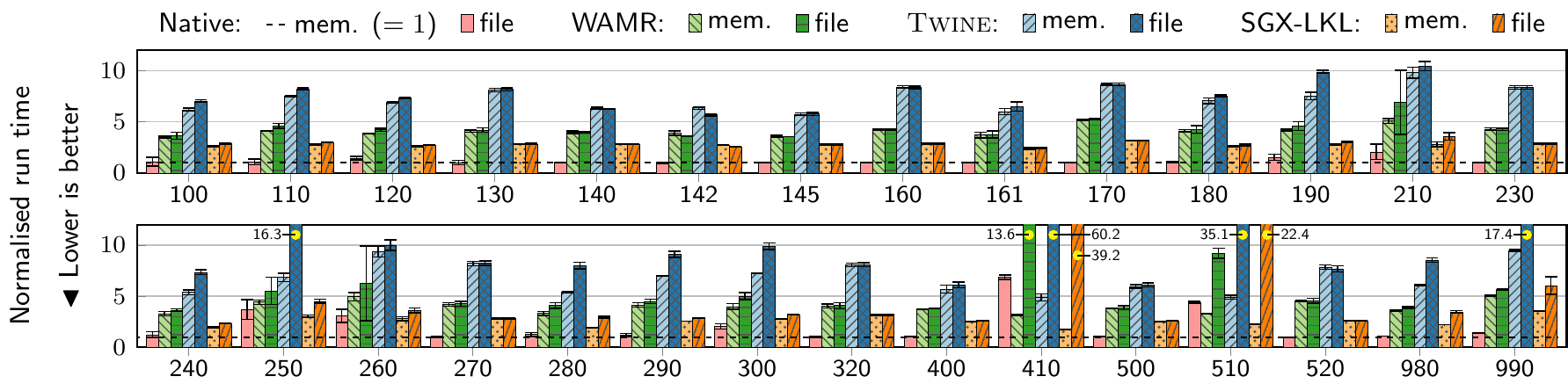}
	\caption{Relative performance of SQLite \texttt{Speedtest1} benchmarks.}
	\label{fig:speedtest}
	\vspace{-12pt}
\end{figure*}

We can split the \polybench test results in 5 groups, based on the proportion between the execution modes (native, WAMR and \sys): \emph{(1)} similar execution time (\texttt{doitgen} and \texttt{seidel-2d}); \emph{(2)} WAMR results similar to each other, but overall slower than to native (\texttt{2mm}, \texttt{3cmm} and \texttt{durbin}); \emph{(3)} \twine is slower than WAMR and native (\texttt{deriche}, \texttt{gemver} and \texttt{lu}); \emph{(4)}
execution times vary significantly between each variant (\texttt{atax}, \texttt{gemm} and \texttt{jacobi-2d}); \emph{(5)} WAMR is faster than its native counterpart.

Wasm applications are usually slower than native ones due to several reasons: increased register pressure, more branch statements, increased code size, \etc.
Following previous work~\cite{DBLP:conf/usenix/JangdaPBG19}, we investigated \texttt{deriche} and \texttt{gramschmidt} using Linux's performance counters, as both produced better results with Wasm (averages over 3 distinct executions).
Our analysis reports 58,002,746 L1 cache misses for native \texttt{deriche} and 57,384,578 for its Wasm counterpart.
Similarly \texttt{gramschmidt} produces 3,679,222,800 and 3,673,458,022 for native and Wasm L1 cache misses. These results confirm that these two Wasm programs produce slightly fewer L1 caching misses (1.1\% and 0.2\%).

We also looked at the impact of memory on performance, given the additional cost for SGX enclaves~\cite{199364}.
Starting from 160\,MiB (the minimum amount to start all of the \polybench tests), we progressively reduced the memory allocated to the Wasm runtime, until the experiments could no longer allocate memory.
We observed that the slowdown in the \texttt{deriche} test is due to hitting the EPC size limit.
Similarly, \texttt{lu} and \texttt{ludcmp} require at least 80\,MiB of memory.

\subsection{SQLite macro-benchmarks}\label{sec:speedtest1}

\begin{figure*}[!htb]
	\centering
	\includegraphics{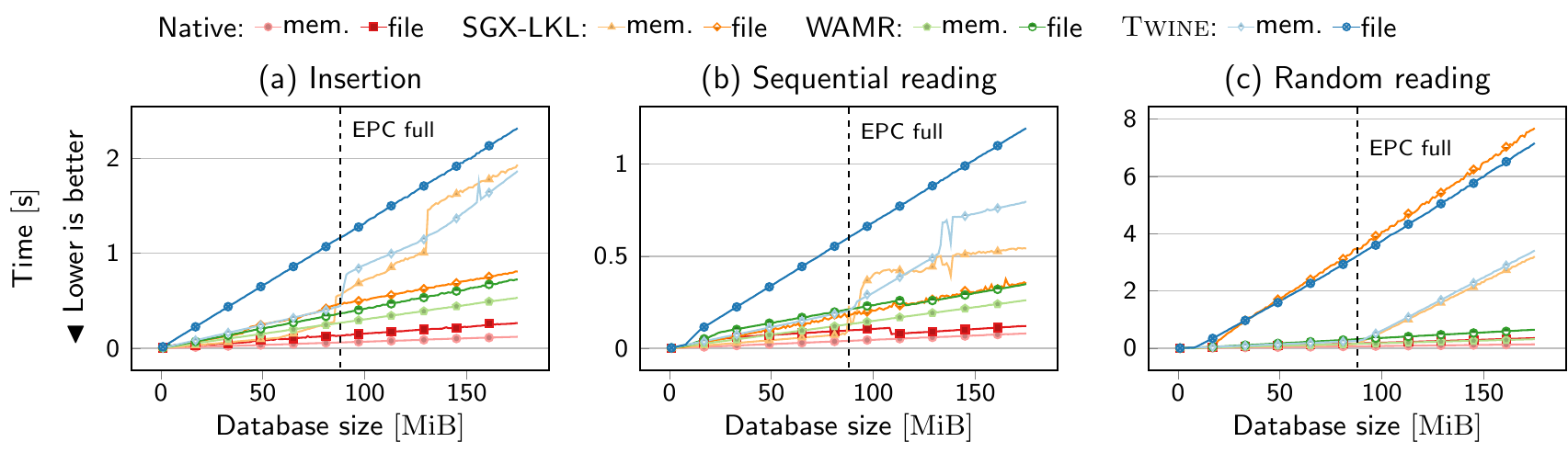}
	\caption{Performance evaluation of SQLite's insertion and reading by technological variants.}
	\label{fig:microbenchmarks}
	\vspace{-12pt}
\end{figure*}

SQLite~\cite{5231398} is a widely-used full-fledged embeddable database.
It is perfectly suited for SGX, thanks to its portability and compact size.
For this reason, we thoroughly evaluated it as a showcase for performance-intensive operations and file system interactions.
SQLite requires many specific OS functions that are missing from the WASI specifications, due to standardisation and portability concerns in Wasm.
Therefore, we relied on SQLite's virtual file system (VFS), and accesses to the file system are translated into the WASI API. Our modified virtual file system implements the minimal requirements to make SQLite process and persist data, reducing the POSIX functions to be supported by \twine WASI layer.
We used one of the official templates (\texttt{test\_demovfs}) to override the OS interface of SQLite since it relies on a few POSIX functions covered by the WASI specifications.
SQLite uses a 2,048-page cache of 4\,KiB each (for a cache size of 8\,MiB) with the default (normal) synchronous mode and the default (delete) journal mode.
Besides, we use an alternate memory allocator  (\texttt{SQLITE\_ENABLE\_MEMSYS3}) to provide a large chunk of pre-allocated memory used for the database instance and records.

Since memory allocation in SGX enclaves is expensive (in some tests, it took up to 45\% of the CPU time to allocate it while inserting records in the database), memory preallocation greatly optimises performance when the database size is known in advance. 
We executed SQLite v3.32.3-amalgamation (\ie, a single-file version of the full SQLite program).
First, we used SQLite's own performance test program, \texttt{Speedtest1}~\cite{speedtest1}, running 29 out of the available 32 tests, covering a large spectrum of scenarios (we excluded 3 experiments because of issues with SQLite VFS).
Each \texttt{Speedtest1} experiment targets a single aspect of the database, \eg, selection using multiple joints, the update of indexed records, \etc.
Tests are composed of an arbitrary number of SQL queries, potentially executed multiple times depending on the load to generate.
\Cref{fig:speedtest} shows our results, normalised against the native execution.
We include results for in-memory configurations as well as for a persisted database, where WASI is used.

While we provide additional details below, we observed across all tests that the WAMR's slowdown relative to native on average is \speedtestWamrMemToNativeRatio$\times$ and \speedtestWamrFileToNativeRatio$\times$ for in-memory and in-file database respectively.
\sys's slowdown relative to WAMR is \speedtestTwineMemToWamrRatio$\times$ and \speedtestTwineFileToWamrRatio$\times$ for in-memory and in-file database.

Experiments 100--120, 180--190, 230, 240, 270--300, 400 and 500 update the database (\eg, creating tables, inserting, updating and deleting records).
They share a similar pattern of performance penalty according to the variants.
Experiments 130, 140 and 145--170 indicate the same performance for in-memory and persistent databases: since they only execute read operations, they act on the page cache, with no file system interaction.
Using SGX with a persistent database adds a considerable overhead under certain circumstances. In particular, experiments 410 and 510, which overflow the page cache and randomly read records, cause additional latency due to the file system interaction, exacerbated by enclave \texttt{OCALL}s and encryption, up to \speedtestExpFourOneZeroTwineMemVsFileRatio$\times$ and \speedtestExpFourOneZeroSgxLklMemVsFileRatio$\times$ for \sys and SGX-LKL respectively compared to the equivalent queries using an in-memory database.
Interestingly, experiments 142 (multiple \texttt{SELECT} with \texttt{ORDER BY}, non-indexed) and 520 (multiple \texttt{SELECT DISTINCT}) show faster results using a persistent database on-file for all the execution modes.
Test 210 is I/O intensive: it alters the database schema and, consequently, all the records.
Similarly, experiment 260 issues a wide-range of \texttt{SELECT} to compute a sum, explaining the high execution time across all execution modes, with a small overhead for SGX variants.
In addition, test 250 is highly I/O intensive with a persisted database, because it updates every record of a table, requiring to reencrypt most of the database file.

Finally, 990 is a particular case of database housekeeping.
It gathers statistics about tables and indices, storing the collected information in internal tables of the database where the query optimiser can access the information and use it to help make better query planning choices.
The longer execution time of \sys and SGX-LKL with a persistent database is explained by the added complexity of I/O from the enclave.

\subsection{Breakdown of SQLite Macro-benchmarks}\label{sec:microbenchmarks}

\newcommand{\twinebetter}{\textsuperscript{*}}
\begin{table}
\small
\setlength{\tabcolsep}{2pt}
\caption{Comparison of the technologies in normalised run time.}
\rowcolors{1}{gray!0}{gray!10}
\begin{tabularx}{\columnwidth}{XS[table-column-width = 10mm]*4{S[table-column-width = 10mm, table-format = 2.1]}}
\toprule
\rowcolor{gray!25}
& &\multicolumn{2}{c}{SGX-LKL} &\multicolumn{2}{c}{\sys}\\[-3pt]
\rowcolor{gray!25}
&{\makecell[c]{\multirow{-2}*{WAMR}}} &{\raisebox{-1px}{\makecell[c]{\scriptsize{$<$EPC}}}} &{\raisebox{-1px}{\makecell[c]{\scriptsize{$\geq$EPC}}}} &{\raisebox{-1px}{\makecell[c]{\scriptsize{$<$EPC}}}} &{\raisebox{-1px}{\makecell[c]{\scriptsize{$\geq$EPC}}}}\\
\midrule
Insert mem.     &\insertWamrMem    &\insertLklMemBepc    &\insertLklMemAepc    &\insertTwineMemBepc                &\insertTwineMemAepc\twinebetter\\
Insert file     &\insertWamrFile   &\insertLklFileBepc   &\insertLklFileAepc   &\insertTwineFileBepc               &\insertTwineFileAepc\\
Seq. read mem.  &\readSeqWamrMem   &\readSeqLklMemBepc   &\readSeqLklMemAepc   &\readSeqTwineMemBepc               &\readSeqTwineMemAepc\\
Seq. read file  &\readSeqWamrFile  &\readSeqLklFileBepc  &\readSeqLklFileAepc  &\readSeqTwineFileBepc              &\readSeqTwineFileAepc\\
Rand. read mem. &\readRandWamrMem  &\readRandLklMemBepc  &\readRandLklMemAepc  &\readRandTwineMemBepc              &\readRandTwineMemAepc\\
Rand. read file &\readRandWamrFile &\readRandLklFileBepc &\readRandLklFileAepc &\readRandTwineFileBepc\twinebetter &\readRandTwineFileAepc\twinebetter\\
\bottomrule
\noalign{\vskip 2pt} 
\multicolumn{2}{l}{\cellcolor{gray!0}\scriptsize{(Native run time = 1)}} & \multicolumn{4}{r}{\cellcolor{gray!0}\scriptsize{\twinebetter\sys is faster than SGX-LKL.}}\\
\end{tabularx}
\label{tbl:microbenchmarks}
\end{table}

To better understand the source of performance penalties observed, we designed a suite of tests for common database queries, including insertion, sequential and random reading (measured separately because of different complexity~\cite{199364}), and inspired by a similar benchmark suite~\cite{Sartakov2018Stanlite}.
The tests use a single-table with an auto-incrementing primary key and a blob column.
For sequential insertions, the blob column is iteratively filled by an array of random data (1\,KiB) using a pseudorandom number generator (PRNG, same as \texttt{Speedtest1}).
Next, records are selected in the order they have been inserted (\texttt{WHERE} clause).
Finally, we selected one random entry at a time.
The database is initialised with 1\,k records (\ie, 1\,MiB in total) and iteratively increases that amount by 1\,k entries at the time, up to 175\,k records (\ie 175\,MiB).
We evaluated 4 variants: a native version of SQLite running either outside or inside of an enclave, and an ahead-of-time Wasm version running either outside or inside of an enclave.
For each of them, we include results for in-memory and on-file databases.
The performance results for \sys (in-file) are based on the enhanced version of IPFS, which reduces the latency of the read/write operations.
The details of the improvement of IPFS are covered in \S\ref{sec:profiling}.
\Cref{tbl:microbenchmarks} summaries the obtained results, where values on each line are normalised with the run time of the native variant.
The run time is the median of the queries' execution time, either from 1\,k to 175\,k records for native and WAMR, or split into two parts for SGX-LKL and \sys, going from 1\,k to EPC size limit and from that limit to 175\,k.

\Cref{fig:microbenchmarks}a shows the results regarding the insertion of records.
While the variants outside the enclave perform steadily, the in-memory variant is affected by the EPC limits.
This is expected due to costly swapping operations~\cite{IntelCorporation2018SGXperf}.
The cost of operations with the persistent database with \sys increases linearly because of the additional file encryptions.
The SGX-LKL implementation has a more optimal approach for inserting sequential elements and follows the trend of \sys's in-memory performance.

\Cref{fig:microbenchmarks}b shows the execution time to sequentially read all records.
The variants outside of the enclave have rather linear costs, with a slight drop when the database is filled with 114\,k records.
We were concentrated in \twine performance, so we did not look into this slightly unexpected behaviour.
It remains to be further investigated later on.
\sys and SGX-LKL with an in-memory database has a sharp increase beyond the EPC size limit due to the enclave paging.
\sys with a database on file performs the best while the database fits in 8\,MiB (\ie, the configured cache of SQLite).
A similar increase is observed up to 16\,MiB (twice the cache size).
To prove this overhead relates to the cache, we increased the cache size to 16\,MiB, noticing the sharp increase stops at 32\,MiB.
We observed similar trends by swapping the WASI layer with the one from WAMR (without any encryption and direct POSIX calls).
Consequently, we identify in the SGX memory accesses the root cause of such performance penalties.

\Cref{fig:microbenchmarks}c depicts the execution time for random readings.
The costs of all the variants increase linearly with the database's size, except for SGX in-memory database variants due to EPC limits.
Random reading triggers the enclave paging mechanism more often because the spatial locality of the requested records is no longer smaller than the size of a single memory page.
Finally, the case of in-file random reading highlights where \sys shines, by providing faster performance compared to SGX-LKL, of \readRandLklVsTwineFileBepc$\times$ before the EPC limit and \readRandLklVsTwineFileAepc$\times$ afterwards.
A similar performance increase is noticed for the in-memory insertion above the EPC limit with a gain of \insertLklVsTwineMemAepc$\times$.

As a result, \sys has slower performance results than SGX-LKL, which is expected due to the overhead incurred by Wasm.
Nonetheless, \sys provides similar but faster operations than SGX-LKL when it involves random access to files and for the insertion in-memory once the EPC threshold is reached while being outperformed for the other use cases.
\Cref{sec:cost-factors} further analyses whether SGX is responsible for this behaviour.

\subsection{Cost factors assessment of SQLite micro-benchmarks}\label{sec:cost-factors}

As part of identifying the performance penalties and bottlenecks introduced by the technologies surrounding SQLite, we performed a comprehensive analysis of the cost factors one can expect by using SGX and Wasm, either independently or in combination.
We identified two aspects of costs:
\emph{(1)}~the time required to build and deploy an application that occurs on the developers' premises, and \emph{(2)}~the time and storage space required to execute an application on the untrusted platform. 

\Cref{tbl:cost-factors}a summarises the time overheads we observed with the SQLite micro-benchmarks (175\,k records).
As different kinds of costs are involved depending on the variant, we do not indicate totals in the table.
The native one is composed of a single executable binary, while SGX-LKL requires the same executable binary and a disk image, which is an abstraction introduced to store the code and data securely.
The two variants that use Wasm require an executable binary and a Wasm artifact containing the SQLite code.
For both variants, we measured the time for AoT compilation as well.
For launching, we measured the time from the process creation to the start of the database initialisation.
The variants without SGX are naturally faster since they do not have to initialise the enclave.
The initialisation of \sys is \launchTimeTwineVsSgx$\times$ faster than SGX-LKL because the enclave is heavier than \sys's and the benchmarks executable is encrypted on the disk image.

\Cref{tbl:cost-factors}b indicates the components' size for the compiled artifacts and other prerequisite software on disk as well as in the resident memory.
The native variant is stored in a single executable binary file.
SGX-LKL has a heavier sized executable and a much larger enclave binary.
The latter contains a generic program that is only loaded once and runs any other program stored in a disk image (in our case, the SQLite benchmarks).
A disk image is necessary for SGX-LKL, which it maps into RAM.
We generated an ext4-formatted file system, whose size is fixed at build time to be big enough to store our SQLite micro-benchmarks programs and results.
\sys have a lightweight runtime, with a reduced memory footprint in the enclave, since the executable binary loaded into the enclave is only SQLite and the benchmarks.
Also, \sys does not need an image file as it relies on the host file system, keeping its content secure thanks to IPFS.
When loaded in RAM (last lines in \Cref{tbl:cost-factors}b), the variants occupy different amounts of memory.
Native and Wasm variants store the database records in the process address space (no enclaves).
\sys and SGX-LKL store records inside their enclaves, resulting in less memory consumed outside.
The enclave sizes were configured to be just big enough to store 175\,k records.

\begin{figure*}[!htb]
	\begin{minipage}[t]{\columnwidth}
		\centering
		\includegraphics{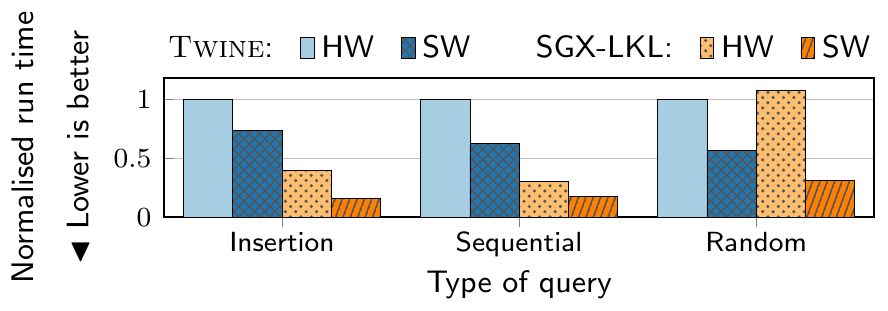}
		\caption{Normalised run time for SGX variants with in-file database.}
		\label{fig:execution-time}
	\end{minipage}
	\hfill
	\begin{minipage}[t]{\columnwidth}
		\centering
		\vspace*{-83px}
		\includegraphics{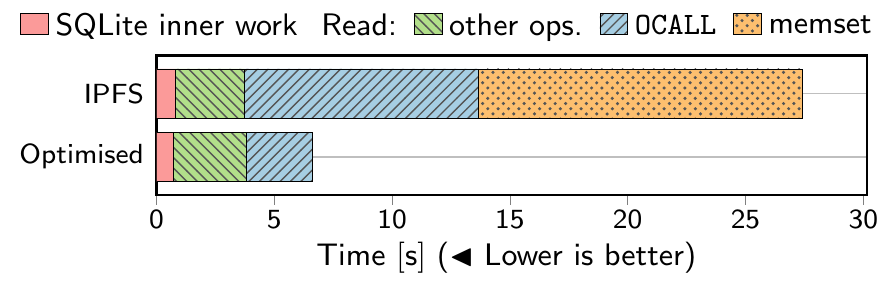}
		\caption{Time breakdown before and after the optimisations.}
		\label{fig:profiling}
	\end{minipage}
	\vspace{-12pt}
  \end{figure*}

Finally, \Cref{fig:execution-time} depicts the overhead incurred by the introduction of SGX in the breakdown of the micro-benchmarks using an in-file database.
In particular, it compares the SGX hardware mode where the SGX memory protection is enabled and the software mode where the SGX protection is emulated.
The normalised run time is the median of the queries' execution time, from 1\,k to 175\,k records compared to \sys hardware mode.
While the insertion and sequential reading time follow a similar trend, the performance of SGX-LKL in hardware mode for the random reading suffers from a slow down.
Since SGX-LKL in software mode does not encounter this issue, the performance loss is assignable to Intel SGX.

\def\labelcolumnlength{28mm}
\begin{table}[t]
\small
\setlength{\tabcolsep}{2pt}
\caption{Cost factors of the micro-benchmarks.}
\rowcolors{1}{gray!10}{gray!0}
\begin{tabularx}{\columnwidth}{p{\labelcolumnlength}RRRR}
\toprule
\rowcolor{gray!25}
\textit{(a)~Times {[ms]}} &Native                        &\textsc{sgx-lkl}                    &WAMR                        &\sys\\
\midrule
Compile runtime           & ---                          &\compilationSgxAverage              &\compilationWasmBinAverage  &\compilationWasmSgxBinAverage\\
Compile Wasm              & ---                          & ---                                &\compilationWasmAverage     &\compilationWasmAverage\\
Compile x86/AoT           &\compilationNativeAverage     &\compilationNativeAverage           &\optimizationWasmAotAverage &\optimizationWasmAotAverage\\
Generate disk image       & ---                          &\optimizationSgxLklDiskImageAverage & ---                        & ---\\
Launch                    &\launchTimeNativeAverage      &\launchTimeSgxAverage               &\launchTimeWasmAverage      &\launchTimeWasmSgxAverage\\
\bottomrule
\rowcolor{gray!25}
\textit{(b)~Sizes {[KiB]}} &Native                       &\textsc{sgx-lkl}                    &WAMR                        &\sys\\
\midrule
Executable, disk           &\sizeNativeAverage           &\sizeSgxLklBinAverage               &\sizeWasmBinAverage         &\sizeWasmSgxBinAverage\\
Enclave, disk              & ---                         &\sizeSgxLklEnclaveAverage           & ---                        &\sizeWasmSgxEnclaveAverage\\
Wasm artifact, disk        & ---                         & ---                                &\sizeWasmAverage            &\sizeWasmAverage\\
AoT artifact, disk         & ---                         & ---                                &\sizeWasmAotAverage         &\sizeWasmAotAverage\\
Disk image                 & ---                         &\sizeSgxLklDiskImageAverage         & ---                        & --- \\
Executable, memory         &\memoryNativeInMemoryAverage &\memorySgxInMemoryAverage           &\memoryWasmInMemoryAverage  &\memoryWasmSgxInMemoryAverage\\
Enclave, memory            & ---                         &\memorySgxEnclaveSize               & ---                        &\memoryWasmSgxEnclaveSize\\
\bottomrule
\end{tabularx}

\rowcolors{1}{gray!10}{gray!0}

\label{tbl:cost-factors}
\end{table}

\subsection{SQLite profiling and optimised IPFS}\label{sec:profiling}

To conclude our experimental analysis, we performed in-depth profiling of the basic SQLite operations, in particular looking at the overheads from the IPFS layer, where we observed the highest slowdowns.
Our analysis allows us to suggest small changes to the SGX SDK that, remarkably, speed up the handling of protected files up to \randReadingRatio$\times$.

We instrumented and profiled the Intel SGX Protected FS. 
It is composed of two modules: one statically linked with the enclave's trusted code and the other one statically linked with the untrusted binary that launches the enclave.
We broke down these two modules into components (\eg, cryptography, node management of the cached data, trusted and untrusted high-level API).
We profiled those with POSIX \texttt{clock} calls.
We manually instrumented the Wasm runtime to profile every implemented WASI functions related to the file system.
The profiling results exclude the execution time to retrieve the current time from the enclave: while this takes 4ms on average, its recurring usage might lead to unexpected profiling results.

We identified the main performance contributions for random reading in the following:
\emph{(1)}~clearing the memory (\texttt{memset}),
\emph{(2)}~\texttt{OCALL}'ing the untrusted functions of the SDK of SGX and call the POSIX functions,
\emph{(3)}~reading the database entries, and
\emph{(4)}~SQL inner works (\ie, cache management).
\Cref{fig:profiling} shows the costs of such operations while randomly reading the records.
The IPFS bar of the breakdown plot indicates that \memsetRatio\% of the time is spent on clearing the memory of the enclave, \ocallRatio\% to transition back and forth (to retrieve the file's content from the untrusted world, \otherOperationsRatio\% for reading operation, and only \sqliteRatio\% are dedicated to SQLite.

IPFS internally manages the content of a protected file by decomposing it into \emph{nodes}.
A node corresponds to a block of data to encrypt/decrypt.
These nodes are stored in a least recently used (LRU) cache. Each node contains two buffers of 4\,KiB each to store the ciphertext and the plaintext.
When a node is added to the cache, its entire data structure is cleared.
Since a SGX memory page is 4\,KiB~\cite{cryptoeprint:2016:086}, at least two pages must be cleared in addition to metadata contained in that structure, such as node identifiers and miscellaneous flags.
When a node is removed, the plaintext buffer is cleared as well, which corresponds to (at least) one SGX memory page.

While initialising structure data members is a good practice in C++ as they are left to indeterminate values by default, it has a significant performance impact in the context of SGX.
The functions that add nodes set several fields after clearing the node structure.
The ciphertext is then read from the untrusted part of the application to be stored in the corresponding buffer and decrypted into the other buffer.
This means the only requirement for initializing the class data members is to set a default value to the fields not assigned.
We propose to remove the clearing operations, replacing them by setting the remaining fields to zero.
Thus, we preserve the initial behaviour of the code, while sparing the valuable time of memory clearing the structure, that is overwritten anyway.
Similarly, upon a node is dropped from the cache, the plaintext buffer is cleared before releasing the node (\ie, using C++'s \texttt{delete}).
While this is a good practice to flush the memory of confidential values when no longer needed, we assume SGX shields the enclave's memory.
Given our considered threat model, no adversary is able to read that, even if sensitive values are left in the SGX memory pages.
For this reason, we also propose to remove the clearing operation for the plaintext in the disposed nodes.

Finally, we look at the time spent reading the file content. 
The function responsible for this task issues an \texttt{OCALL}, crossing the secure enclave boundary to read the content of the database file.
Our profiling measures show that while the untrusted POSIX calls are fast, a bottleneck exists in the code generated by the SGX tool \texttt{edger8r} to interface the untrusted part of the application with the enclave. 
The \texttt{edger8r} tool facilitates the development of SGX enclaves generating edge routines to interface untrusted application and the enclave and enabling one to issue \texttt{ECALL}s and \texttt{OCALL}s straightforwardly.
The edge functions responsible for reading the files outside of the enclave specifies that the buffer that contains the data must be copied from the untrusted application into the enclave secure memory.
IPFS decrypts it after issuing the \texttt{OCALL} and stores the plaintext into a buffer of the node structure.
Our profiling indicates that \copyRatio\% of the time is spent to complete this ciphertext copy from the untrusted application.
We propose to remove this copy to the enclave altogether.
Instead, we provide a pointer to the buffer located in the untrusted memory to the enclave, from where the library directly decrypts.
With the new implementation, an adversary may attempt to perform a timing attack to alter the ciphertext between the authentication of the data and its decryption, as the authenticated mode of operation of AES-GCM is \emph{encrypt-then-MAC}.
We suggest using a different encryption algorithm in this case, such as AES-CCM~\cite{aesccm}, that calculates the MAC from plaintext instead (\emph{MAC-then-encrypt}).
The cryptography libraries available in Intel's SGX SDK already includes this cipher.
With AES-CCM, the authentication is verified based on data already securely stored in the enclave.
The cost for decrypting a block that happens to fail authentication is small compared to a systematic copy of the buffer and remains a rare event when used legitimately.

The performance gains of our optimised IPFS can be seen in \Cref{fig:profiling} for random reading queries with 175\,k records.
The time for clearing the memory has now been eliminated, and the file reading operations represent \copyInvertedRatio\% of the initial execution time.
Compared to Intel's version, insertion achieves a \insertRatio$\times$ speedup and \seqReadingRatio$\times$ for sequential reading. 
Finally, for random reading, we achieved a \randReadingRatio$\times$ speedup.

\section{Conclusion}\label{sec:concl}

The lack of trust when outsourcing computation to remote parties is a major impediment to the adoption of distributed architectures for sensitive applications.
Whereas this problem has been extensively studied in the context of cloud computing across large data centres, it has been only scarcely addressed for decentralised and resource-constrained environments as found in IoT or edge computing.
In this paper, we proposed an approach for executing unmodified programs in WebAssembly (Wasm)---a target binary format for applications written in languages supported by LLVM, such as C, C++, Rust, Fortran, Haskell, \etc---within lightweight trusted execution environments that can be straightforwardly deployed across client and edge computers.
\twine is our trusted runtime with support for execution of unmodified Wasm binaries within SGX enclaves.
We provide an adaptation layer between the standard Wasm system interface (WASI) used by the applications and the underlying OS, dynamically translating the WASI operations into equivalent native system calls or functions from secure libraries purposely built for SGX enclaves.
Our in-depth evaluation shows performance on par with other state-of-the-art approaches while offering strong security guarantees and full compatibility with standard Wasm applications.
\twine is freely available as open-source.

 \section*{Acknowledgments}

This publication incorporates results from the VEDLIoT project, which received funding from the European Union’s Horizon 2020 research and innovation programme under grant agreement No 957197.

{\small
\bibliographystyle{IEEEtran}
\bibliography{paper}
}
\end{document}